\documentclass[%
nofootinbib,
amsmath,amssymb,
aps,
prd,
twocolumn,
longbibliography,
superscriptaddress]{revtex4-2}

\usepackage{dsfont}
\usepackage{graphicx}
\usepackage{dcolumn}
\usepackage{bm}
\usepackage[%
  colorlinks=true,
  urlcolor=blue,
  linkcolor=blue,
  citecolor=blue,
]{hyperref}
\usepackage{footnote}
\usepackage{float}
\usepackage{caption}
\usepackage{xcolor}
\usepackage{subfig}
\usepackage[table,xcdraw]{}

\begin{document}

\title{Transport coefficients of chiral fluid dynamics using low-energy effective
models}

\author{Pedro {\sc Nogarolli}}
\email{pedro.nogarolli@cern.ch}
 \affiliation{
 Instituto de F\'\i sica, Universidade Federal do Rio de Janeiro,\\
 CEP 21941-909 Rio de Janeiro, RJ, Brazil 
}

\author{Gabriel S. {\sc Denicol}}
 \email{gsdenicol@id.uff.br}
 \affiliation{
 Instituto de F\'\i sica, Universidade Federal Fluminense,\\
 CEP 24210-346 Niterói, RJ, Brazil 
}

\author{Eduardo S. {\sc Fraga}}
 \email{fraga@if.ufrj.br} 
 \affiliation{
 Instituto de F\'\i sica, Universidade Federal do Rio de Janeiro,\\
 CEP 21941-909 Rio de Janeiro, RJ, Brazil 
}

\begin{abstract}

We investigate the first-order transport coefficients of a fluid made of quasiparticles with a temperature-dependent mass extracted from chiral models. We describe this system using an effective kinetic theory, given by the relativistic Boltzmann equation coupled to a temperature-dependent background field determined from the thermal masses. We then simplify the collision term using the relaxation time approximation and implement a Chapman-Enskog expansion to calculate all first-order transport coefficients. In particular, we compute the bulk and shear viscosities using thermal masses extracted from the linear sigma model coupled with constituent quarks and the NJL model.

\end{abstract}

\maketitle

\section{Introduction}

Heavy ion collisions experiments at the Relativistic Heavy-Ion Collider and the Large Hadron Collider (LHC) provide access to hot and dense deconfined matter, the quark-gluon plasma (QGP). In these experiments, a large elliptic flow has been measured \cite{PhysRevLett.92.112301,gavin_abdel-aziz_2007} and is interpreted as a sign of hydrodynamic behavior. Relativistic fluid-dynamical theories
have been crucial to investigate the properties of the quark-gluon plasma created in ultrarelativistic heavy-ion collisions \cite{Gale:2013da}.
Ideal hydrodynamics was first used to describe the evolution of the QGP and had a good agreement with experimental data \cite{kolb2003hydrodynamic}. This nearly perfect fluid behavior implies a very small shear viscosity, considerably smaller than extrapolations \cite{PhysRevLett.97.152303} based on weak-coupling calculations \cite{PeterArnold_2000}. Recent phenomenological fluid-dynamical models obtained bulk viscosity values \cite{PhysRevC.103.054909,PARKKILA2022137485,PhysRevC.103.054904} that are orders of magnitude larger than predictions derived in the weak-coupling regime \cite{PhysRevD.74.085021}.

The surprising success of fluid-dynamical models in describing the dynamics of the hot and dense matter produced in heavy ion experiments has motivated a broad discussion on how these theories emerge from a microscopic theory. This issue has been extensively addressed in the context of kinetic theory, for relativistic dilute gases \cite{Denicol-rischke_2022}. However, in this case one often neglects the effects of phase transitions, which cannot be trivially included in a kinetic framework. On the other hand, such effects can be considered in kinetic theory when deriving chiral fluid dynamics \cite{Mishustin:1998yc,Dumitru:2000ai,Paech:2003fe,Aguiar:2003pp,Paech:2005cx,Nahrgang:2011vn,Nahrgang:2011mg,Herold:2013bi,Mishustin:2014mka,Grossi:2020ezz,Bluhm:2020mpc,Weickgenannt:2023nge}, which are fluid-dynamical
theories applicable in phase transitions due to the breakdown of chiral symmetry.

The use of chiral effective models represents a simple way to incorporate effects from chiral symmetry, and its breakdown, in strong interactions. In fact, in a theory with spontaneous symmetry breaking, the presence of a condensate will modify the masses, bringing corrections that are functions of the chiral condensate, which is a medium-dependent quantity. Here we adopt the two most commonly used chiral models, the linear sigma model coupled with constituent quarks (LSMq) and the Nambu-Jona-Lasinio (NJL) model, at mean-field level in the framework of finite-temperature field theory \cite{kapusta_gale_2006}. Usually the gas of quarks is treated as a thermal bath in which the long-wavelength modes of the chiral fields evolve, the latter playing the role of an order parameter in a Landau-Ginzburg approach. The procedure that is usually adopted is integrating over the fermionic degrees of freedom, using a classical approximation for the chiral field, to obtain a formal expression for the thermodynamic potential from which one can compute all the physical quantities of interest. The fermionic determinant is usually calculated to one-loop order assuming a homogeneous and static background field \cite{kapusta_gale_2006}. Here, as discussed above, we follow a different path, and resort to kinetic theory. 

In this paper we investigate transport coefficients within chiral fluid dynamics using low-energy effective models to describe the role played by the chiral condensate in the thermal mass that will serve as an input to a kinetic theory description of the quark fluid. The effective thermal masses can be treated as a temperature-dependent background field in the corresponding relativistic Boltzmann equation. We then use a quasiparticle model of the relativistic Boltzmann equation \cite{Rocha:2022fqz} combined with the Chapman-Enskog expansion. In contrast to Ref.~\cite{Chakraborty:2016ttq}, this procedure is implemented with using a recent extension of the relaxation time approximation of the linear collision kernel \cite{Rocha:2021zcw} that is consistent with fundamental conservation laws. In this setting, the transport coefficients are written solely in terms of the temperature-dependent mass and the parameter determining the energy dependence of the relaxation time. We compute the bulk and shear viscosities, the speed of sound, and several thermodynamic quantities, such as the equilibrium energy density, pressure, entropy and trace anomaly. 

The main novelty of our results come from a consistent implementation of the relaxation time approximation. In several previous works (not necessarily employing thermal masses), the Anderson-Witting relaxation time approximation was employed to simplify the collision kernel and obtain simple expressions for transport coefficients. However, this approximation is not consistent with the energy-momentum conservation law, violating it when the relaxation time exhibits an energy dependence or when different matching conditions are employed to define the temperature \cite{Rocha:2021zcw}. In this work, we thus provide simple expressions for the transport coefficients without violating basic conservation laws. This self-consistent framework allowed us to calculate the transport coefficients for different energy dependencies of the relaxation time and also matching conditions. 

The paper is organized as follows. Section \ref{chiral} briefly reviews the LSMq and the NJL model, as well as their respective effective quark masses as a function of the temperature. In Sec. \ref{quasiparticlesection} we summarize the quasiparticle model of the relativistic Boltzmann equation and how to use it to compute thermodynamic quantities. Section \ref{relaxationsection} discusses an improved relaxation time approximation to the linear collision kernel with an energy-dependent relaxation time. Section \ref{chapmanexpansionsection} shows the first-order solution of the Chapman-Enskog expansion of the relativistic Boltzmann equation with a thermal mass and uses this solution to express the transport coefficients. We discuss the results in Sec. \ref{discussionsection} and present our summary and outlook in Sec. \ref{conclusion}. We use natural units $\hbar=k_{b}=c=1$ and a metric signature given by $g^{\mu\nu}={\rm diag}(+,-,-,-)$.

\section{Chiral Effective Models}
\label{chiral}

In this paper, we use two well-known low-energy effective chiral models as ingredients to our chiral fluid dynamics, the LSMq and the NJL model. They provide the medium-dependent effective mass that will will serve as an input in the following kinetic theory description we provide to compute transport coefficients.

The LSMq, also known as the quark-meson model, was shown to be very suitable for the study of the chiral transition \cite{lee-book}. As argued, originally, in Ref. \cite{Pisarski:1983ms}, QCD with two flavors of massless quarks belongs to the same universality class as the $O(4)$ LSM, exhibiting the same qualitative behavior at criticality (see, however, Ref. \cite{Fejos:2024bgl} for a different perspective). The LSMq has also the good feature of being renormalizable \cite{Lee:1968da}, even though this is not necessary for an effective model, and reproduces correctly the phenomenology of strong interactions at low energies. Since its proposal \cite{Gell-Mann:1960mvl}, the LSM has been investigated in different contexts, from the low-energy nuclear theory of nucleon-meson interactions to ultrarelativistic high-energy heavy-ion collisions. Since very early, it has been mostly used to mimic and study the chiral transition of thermal QCD \cite{Baym:1977qb,Bochkarev:1995gi,Bilic:1997sh,Petropoulos:1998gt,Scavenius:1999zc,Roder:2003uz,Scavenius:2000qd,Scavenius:2000bb,Fraga:2004hp,Koide:2004yn,Sasaki:2007qh,Marko:2010cd,Nahrgang:2011mv}, but also to investigate this transition at finite density \cite{Palhares:2010be,Kahara:2010wh,Kroff:2014qxa}

The Lagrangian of the LSMq has the following form
\begin{eqnarray}
{\cal L} &=&
 \overline{\psi}_f \left[i\gamma ^{\mu}\partial _{\mu} - m_f - g(\sigma +i\gamma _{5}
 \vec{\tau} \cdot \vec{\pi} )\right]\psi_f \nonumber\\
&+& \frac{1}{2}(\partial _{\mu}\sigma \partial ^{\mu}\sigma + \partial _{\mu}
\vec{\pi}  \cdot \partial ^{\mu}\vec{\pi} )
- U(\sigma ,\vec{\pi})\;,
\label{lagrangian}
\end{eqnarray}
where
\begin{equation} 
U(\sigma ,\vec{\pi})=\frac{\lambda}{4}(\sigma^{2}+\vec{\pi}^{2} -
{\it v}^2)^2-h\sigma
\label{bare_potential}
\end{equation}
is the self-interaction potential for the mesons, exhibiting both spontaneous and explicit breaking of chiral symmetry. The $N_f=2$ massive fermion fields $\psi_f$ represent the up and down constituent-quark fields $\psi_{f}=(u,d)$. For simplicity, we attribute the same mass, $m_f=m_q$, to both quarks. The scalar field $\sigma$ plays the role of an approximate order parameter for the chiral transition, being an exact order parameter for massless quarks and pions. The latter are represented by the pseudoscalar field $\vec{\pi}=(\pi_{1},\pi_{2},\pi_{3})$. It is customary to group together these meson fields into a $O(4)$ chiral field $\phi =(\sigma,\vec{\pi})$. For simplicity, we discard the pion dynamics from the outset, knowing that they do not affect appreciably the phase conversion process \cite{Scavenius:2000bb}, and focus our discussion on the quark-sigma sector 
of the theory. Nevertheless, pion vacuum properties will be needed to fix the parameters of the Lagrangian in order to reproduce correctly the phenomenology of QCD at low energies and in the vacuum, such as the spontaneous (and small explicit) breaking of chiral symmetry and experimentally measured meson masses. 

The parameters of the Lagrangian are chosen such that chiral $SU_{L}(2) \times SU_{R}(2)$ symmetry is spontaneously broken in the vacuum.  The vacuum expectation values of the condensates are $\langle\sigma\rangle ={\it f}_{\pi}$ and $\langle\vec{\pi}\rangle =0$, where ${\it f}_{\pi}=93$~MeV 
is the pion decay constant. The explicit symmetry breaking term is due to the finite current-quark masses and is determined by the so called Partial Conserved Axial Current hypothesis (PCAC) relation which gives $h=f_{\pi}m_{\pi}^{2}$, where $m_{\pi}=138$~MeV is the pion mass. This leads to $v^{2}=f^{2}_{\pi}-{m^{2}_{\pi}}/{\lambda ^{2}}$.  The value of $\lambda^2 = 20$ leads to a $\sigma$ mass, $m^2_\sigma=2 \lambda^{2}f^{2}_{\pi}+m^{2}_{\pi}$, equal to 600~MeV. In mean field theory, the purely bosonic part of this Lagrangian exhibits a second-order phase transition~\cite{Pisarski:1983ms} at $T_c=\sqrt{2}v$ if the explicit symmetry breaking term, $h$, is dropped.  For $h\ne 0$, the transition becomes a smooth crossover from the restored to broken symmetry phases. For sufficiently small values of $g$ one still finds a smooth transition between the two phases, whereas at larger values the effective potential exhibits a first-order phase transition \cite{Scavenius_2001}. We fix $g=3.3$, so that we have a crossover that is compatible with lattice QCD data.

Since chiral symmetry is spontaneously broken in the vacuum, it is convenient to expand the $\sigma$ field around the condensate, writing $\sigma(\vec{x},\tau) =\langle\sigma\rangle + \xi (\vec{x},\tau)$, where $\tau$ is the imaginary time in the Matsubara finite-temperature formalism. Given the shift above, the fluctuation field $\xi$ is such that $\langle\xi\rangle =0$ and $\xi (\vec{p}=0,\omega=0)=0$. From the phenomenology, one expects that $\langle\sigma\rangle (T \rightarrow \infty) \approx 0$ (being equal in the case of massless quarks). Keeping terms up to $O(\xi^2)$, one obtains the following effective Lagrangian
\begin{eqnarray}
{\cal L}  &=&
\overline{\psi}_f \left[i\gamma ^{\mu}\partial _{\mu} - 
M_q - g\xi \right]\psi_f \nonumber \\
&+& \frac{1}{2}\partial _{\mu}\xi \partial ^{\mu}\xi 
-\frac{1}{2}M_\sigma^2\xi^2 - U(\langle\sigma\rangle)\;,
\label{effective_lagrangian}
\end{eqnarray}
where we have defined the temperature-dependent effective masses
\begin{equation}
M_q \equiv m_q + g \langle\sigma\rangle \;\;,\;\; 
M_\sigma^2\equiv 3\langle\sigma\rangle^2 -\lambda {\it v}^2 \; .
\label{masses}
\end{equation}
Linear terms in $\xi$ can be dropped in the action because of the condition $\xi (\vec{p}=0,\omega=0)=0$. 

\begin{figure}[h]
    \centering
      \includegraphics[width=.48\textwidth]{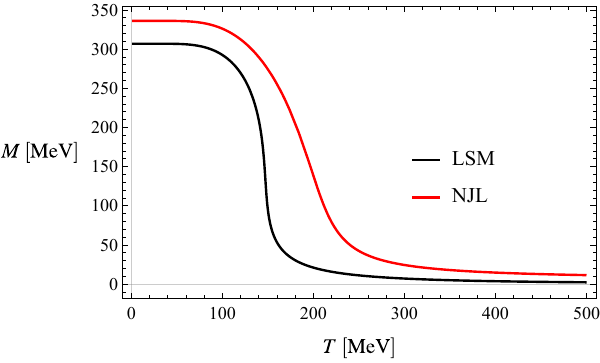}   
\caption{Thermal effective quark mass as a function of the temperature for the LSMq and NJL models.}
\label{massnjl}
\end{figure}

The NJL model is a low-energy proxy of QCD that is widely used in the description of hadron properties and the chiral phase transition \cite{Vogl:1991qt,Klevansky:1992qe,Buballa:2003qv}. Its simplest version includes only scalar and pseudoscalar four-fermion interaction terms and is given by the following Lagrangian \cite{Scavenius:2000qd}:
\begin{equation}
\label{lagrangiannjl}
\mathcal{L}=\bar{q}(i\gamma^{\mu}\partial_{\mu}-m_{0})q+\frac{G}{2}\left[(\bar{q}q)^2+(\bar{q}i\gamma^{5}\vec{\tau}q)^2\right],
\end{equation}
where $m_{0}$ is the small current constituent quark mass. The model is nonrenormalizable since the coupling constant has dimension of $(\textrm{energy})^{-2}$. This brings the necessity to introduce a cutoff $\Lambda$, which is a new parameter of the model. As before, the free parameters are fixed to reproduce the vacuum properties, such as the pion decay constant and pion mass. This yields $G=5.496$ $\textrm{GeV}^{-2}$, $m_{0}=5.5$ MeV and $\Lambda=631$ MeV. 

The constituent quark masses from both models were computed using standard methods in thermal field theory (see, e.g., \cite{Scavenius:2000qd,Scavenius:2000bb}). The for of their temperature dependence is shown in Fig. \ref{massnjl}. Notice that the NJL model and LSMq exhibit a very similar qualitative behavior for their thermal effective masses. However, LSMq has a faster chiral restoration as a function of the temperature as compared to the NJL model. This difference may seem small here, but it has significant qualitative effects in the transport coefficients as will be discussed later.

\section{Quasiparticle model for the Boltzmann equation}
\label{quasiparticlesection}

The relativistic Boltzmann equation is a nonlinear integro-differential equation describing the evolution of the single-particle momentum distribution, $f(x,{\mathbf{p})} \equiv f_{\mathbf{p}}$. 
In this paper, we consider a temperature-dependent mass $M(T)$, so that the Boltzmann equation can be written as \cite{Jeon_1996, DEBBASCH20091818}
\begin{equation}
p^{\mu}\partial_{\mu}f_{\textbf{p}}+\frac{1}{2}\partial_{i}M^{2}(T)\frac{\partial f_{\textbf{p}}}{\partial\textbf{p}_{i}}=C[f_{\textbf{p}}] \, ,
\end{equation}
where $C[f_{\textbf{p}}]$ is the collision kernel. If we assume a one-component gas with classical particles interacting through elastic scattering, $C[f_{\textbf{p}}]$ is given by
\begin{equation}
C[f_{\textbf{p}}]= \int dQdQ'dP' W_{pp'\rightarrow qq'}(f_{\textbf{k}}f_{\textbf{k}'}-f_{\textbf{p}}f_{\textbf{p}'}) \, ,
\end{equation}
where $W_{pp'\rightarrow qq'}$ is the transition rate and we defined the integral
\begin{equation}
\int dP \equiv \int \frac{d\textbf{p}^{3}}{(2\pi)^{3}E_{\textbf{p}}}\, .
\end{equation}
Here $E_{\textbf{p}}^{2}=M^{2}+p^{2}$ is the energy of the particles. 

In what follows, we also use the definition
\begin{equation}
\langle \cdot \cdot \cdot \rangle \equiv \int dP (\cdot \cdot \cdot)f_{\mathbf{p}} \, ,
\end{equation}
and the shorthand notation
\begin{equation}
\partial^{i}_{\mathbf{p}}\equiv \frac{\partial}{\partial \mathbf{p}_{i}}\, .
\end{equation}
%

\begin{figure}[t]
    \includegraphics[width=.48\textwidth]{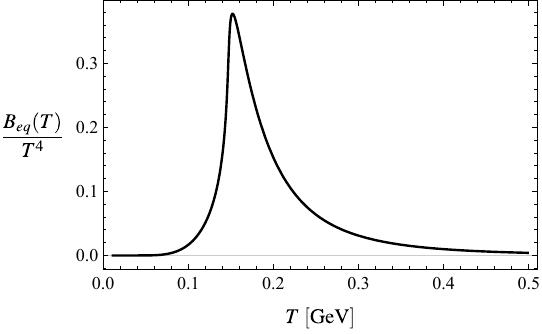}
    \caption{Background field $B$ as a function of the temperature for the LSMq.}
    \label{Bsigma}
\end{figure}
\begin{figure}[t]
    \includegraphics[width=.48\textwidth]{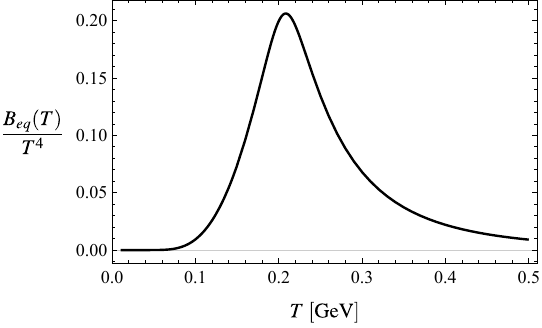}
    \caption{Background field $B$ as a function of temperature for the NJL model.}
    \label{Bnjlfigure}
\end{figure}

Since the temperature in principle varies in space-time, the thermal mass will become a local quantity and, in order to preserve energy-momentum conservation, it becomes necessary to introduce a temperature-dependent background field $B$ \cite{Romatschke_2012, Alqahtani_2015, Jeon_1996, PhysRevD.52.5206} in the energy-momentum tensor. In practice, we have that,
\begin{equation}
\label{energymomentum}
T^{\mu \nu}= \langle p^{\mu}p^{\nu}\rangle+ g^{\mu\nu}B \, ,
\end{equation}
and the conservation of energy-momentum imposes the following relation:
\begin{equation}
\label{background}
\partial_{\mu}B=-\frac{1}{2}\partial_{\mu}M^{2}\langle 1 \rangle.
\end{equation}

The definition of the equilibrium hydrodynamic variables, $T$ and $u^\mu$, is arbitrary and is traditionally implemented via matching conditions \cite{Denicol-rischke_2022}. In this paper, we choose the following matching condition:
\begin{equation}
\label{matching1}
\int dP f_{\mathbf{p}}=\int dP f_{\mathbf{p}}^{(0)} \,,
\end{equation}
where $f^{(0)}$ is the equilibrium distribution function
\begin{equation}
f^{(0)}_{\mathbf{p}}\equiv\exp(-\beta u^{\mu}p_{\mu}) \,.
\end{equation}
Above, $\beta=1/T$ is the inverse temperature and $u^\mu$ is the four-velocity.
The matching condition above defines the temperature $T=1/\beta$. Using the matching condition (\ref{matching1}), the differential equation for the background field $B$, Eq. (\ref{background}), becomes
\begin{equation}
\partial_{\mu}B=-\frac{1}{2}\partial_{\mu}M^{2}\langle 1 \rangle_{0} \,,
\end{equation}
where the notation $\langle \cdot \cdot \cdot \rangle_{0}$ denotes momentum integrals over the local equilibrium distribution function. The definition of the four-velocity is obtained by imposing that the energy diffusion four-current vanishes, as is usually done in simulations of heavy ion collisions.

\begin{figure}[t]
\includegraphics[width=.4\textwidth]{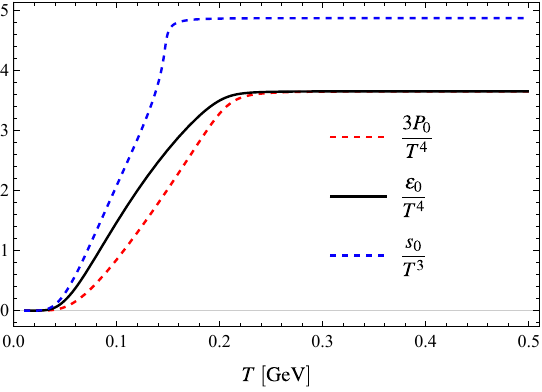}
    \caption{Pressure, energy density and entropy as functions of the temperature for LSMq.}
    \label{epfigure}
\end{figure}

\begin{figure}[t]
\includegraphics[width=.4\textwidth]{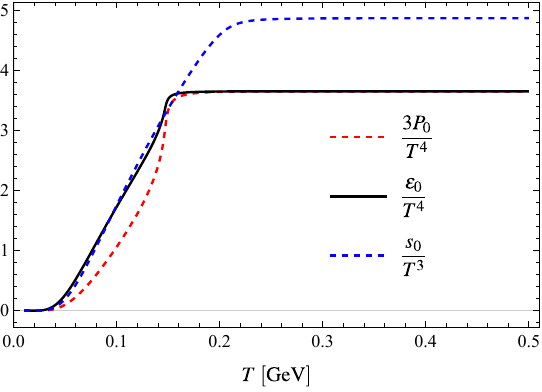}
    \caption{Pressure, energy density and entropy as functions of the temperature for the NJL model.}
    \label{epnjlfigure}
\end{figure}

 \begin{figure}[h]
   \centering
    \includegraphics[width=.48\textwidth]{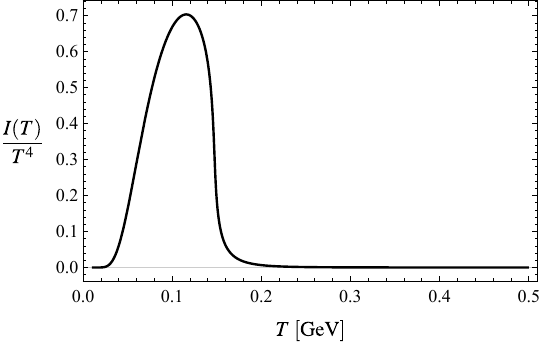}
    \caption{Trace anomaly as a function of the temperature for the LSMq.}
    \label{tracefigure}
\end{figure}

\begin{figure}[h]
   \centering
    \includegraphics[width=.48\textwidth]{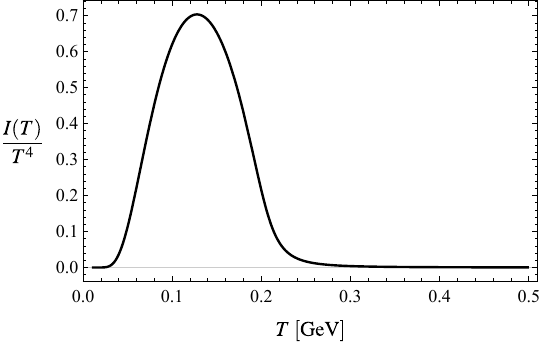}
    \caption{Trace anomaly as a function of the temperature for the NJL model.}
    \label{tracenjlfigure}
\end{figure}

For this matching condition, we can assume that $B$ only depends on the temperature, and hence obtain the following simple relation,
\begin{equation}
\label{bt}
\frac{\partial{B}}{\partial{T}}=-\frac{1}{2}\frac{\partial{M^{2}}}{\partial{T}}\langle 1 \rangle_{0} \,.
\end{equation}
Calculating the momentum integral of Eq. (\ref{bt}) results in the following identity:
\begin{equation}
\langle 1 \rangle _{0}= \frac{\nu_{q}TM}{2\pi^{2}}K_{1}(w) \,,
\end{equation}
where $w=M(T)/T$, $\nu_{q}$ is the degeneracy factor, and $K_{n}(w)$ is the $n$th Bessel Function of the second kind. Using this, integrating Eq. (\ref{bt}) over the temperature and imposing the boundary condition $B(0)=0$, we obtain
%
%
\begin{equation}
B(T)=-\frac{\nu_{q}}{2\pi^{2}}\int_{0}^{T}dx \, M^{2}(x)\,x\,K_{1}\left(\frac{M(x)}{x}\right)\partial_{x}M(x)\, .
\end{equation}

We show the behavior of the background field $B$ as a function of the temperature using the thermal mass provided by the LSMq and the NJL model in Figs.~\ref{Bsigma} and \ref{Bnjlfigure}, respectively. The curves are qualitatively and quantitatively very similar, as expected, with the LSMq exhibiting a sharper and more pronounced peak.

After computing the temperature-dependent background field $B$, one can obtain several thermodynamic quantities from the energy-momentum tensor, Eq. (\ref{energymomentum}). The equilibrium energy density, pressure and entropy density are computed using the thermal mass from the LSMq and NJL models and are shown in Figs. \ref{epfigure} and \ref{epnjlfigure}, respectively.

The trace anomaly, $I_{0}=\epsilon_{0}-3P_{0}$, computed using the thermal masses from the LSMq and NJL models are shown in Figs. \ref{tracefigure} and \ref{tracenjlfigure}, respectively. In both cases, $I_{0}$ goes to zero and reaches the limit of a relativistic ideal gas, i.e., $P_{0}=\frac{1}{3}\epsilon_{0}$, after the critical temperature. In fact, after the approximate chiral restoration the thermal mass becomes very small and we reach the limit of a relativistic Boltzmann equation with massless particles in the absence of external forces, which gives $P_{0}=\frac{1}{3}\epsilon_{0}$.

\begin{figure}[h]
     \centering
     \includegraphics[width=.48\textwidth]{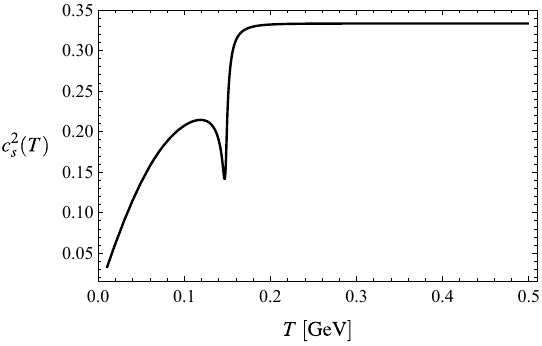}
    \caption{Speed of sound squared as a function of the temperature for the LSMq.}
    \label{csmodel}
    \end{figure}

    \begin{figure}[h]
     \centering
     \includegraphics[width=.48\textwidth]{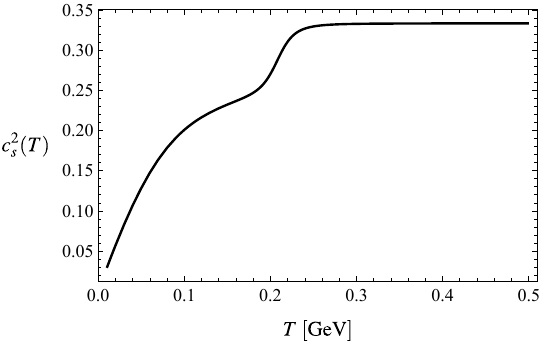}
    \caption{Speed of sound squared as a function of the temperature for the NJL model.}
    \label{csnjlmodel}
    \end{figure}
   
In Figs. \ref{csmodel} and \ref{csnjlmodel} we show the speed of sound computed using the thermal masses from the LSMq and NJL models, respectively. One can see that the major qualitative difference between these chiral effective models is that the speed of sound in the LSMq decreases abruptly right before the critical temperature is reached, whereas in the NJL model $c_s^2$ exhibits a rather smooth behavior. This is due to chiral restoration occurring within a shorter range of temperatures in the LSMq as compared to the NJL model.

\section{Relaxation Time Approximation}
\label{relaxationsection}

The relaxation time approximation consists in inserting a phenomenological time scale within which the system reaches equilibrium. The approximation proposed in Ref. \cite{Anderson:1974nyl} is given by
\begin{equation}
\hat{L}\phi_{\mathbf{p}}\approx\hat{L}_{\rm RTA}\phi_{\mathbf{p}}=-\frac{E_{\mathbf{p}}}{\tau_{R}}f_{\mathbf{p}}^{(0)}\phi_{\mathbf{p}} \,,
\end{equation}
where $\tau_{R}$ is the relaxation time and
\begin{equation}
\phi_{\mathbf{p}}=\frac{f_{\mathbf{p}}-f^{(0)}_{\mathbf{p}}}{f^{(0)}_{\mathbf{p}}} \, .  
\end{equation}
However, this approach has problems, the main issue being that this approximation does not respect the conservation of number of particles and energy momentum. One option is to impose the Landau matching condition to force energy-momentum conservation. This is still not general, since it is only valid if the relaxation time has no momentum dependence. 

Here, we adopt a formulation that circumvents this issue \cite{Rocha:2021zcw}. The relaxation time approximation is defined as
\begin{equation}
\label{RTAbrabo}
\hat{L}_{\rm RTA}\propto\mathds{1}-\sum_{n=1}^{5}|\lambda\rangle\langle\lambda|.
\end{equation}
The identity operator corresponds to the usual approximation \cite{Anderson:1974nyl} and $|\lambda_{n}\rangle$ are the five degenerate orthonormal eigenvectors of $\hat{L}$ that have vanishing eigenvalues, i.e., $\hat{L}|\lambda_{n}\rangle=0$. This guarantees that $\hat{L}_{\rm RTA}|\lambda_{n}\rangle=0$, independently of any matching condition or energy dependence of the relaxation time. For this paper, the vanishing chemical potential implies that the only microscopic conserved quantities are energy and momentum. Therefore, we can remove the counterterm associated with particle number conservation that was originally in \cite{Rocha:2021zcw}, obtaining
\begin{equation}
\begin{aligned}
\label{rtatudo}
\hat{L}_{RTA}\phi_{\mathbf{p}}
=-\frac{E_{\mathbf{p}}}{\tau_{R}}
&f_{\mathbf{0p}}\Biggl [\phi_{\mathbf{p}}^{(1)}-
\frac{\left\langle\frac{\phi_{\mathbf{p}}^{(1)}E_{\mathbf{p}}^{2}}{\tau_{R}}\right\rangle_{0}}{\left\langle\frac{E_{\mathbf{p}}^{3}}{\tau_{R}}\right\rangle_{0}}E_{\mathbf{p}}
\\
&
-\frac{\left\langle\frac{\phi_{\mathbf{p}}^{(1)}E_{\mathbf{p}}p^{\langle\mu\rangle}}{\tau_{R}}\right\rangle_{0}}{\frac{1}{3}\left\langle\Delta^{\alpha\beta}p_{\alpha}p_{\beta}\frac{E_{\mathbf{p}}}{\tau_{R}}\right\rangle_{0}}p_{\langle\mu\rangle}\Biggl] \,,
\end{aligned}
\end{equation}
where we defined the irreducible moment $p_{\langle\mu\rangle}=\Delta_{\mu\nu}p^{\nu}$, which makes use of the projection tensor $\Delta_{\mu\nu}=g_{\mu\nu}-u_{\mu}u_{\nu}$, where $u_{\mu}$ is the four-velocity of the fluid.

The energy dependence of the relaxation time is parametrized as
\begin{equation}
\tau_{R}=t_{R}\left(\frac{E_{\mathbf{p}}}{T}\right)^{\gamma},
\end{equation}
where $t_{R}$ has no energy dependence. Not only is the improved relaxation time approximation (RTA) proposed in Ref. \cite{Rocha:2021zcw} consistent with the five microscopic conserved quantities, but it is also more general than imposing the Landau frame with an energy-independent relaxation time. 

Here, $\gamma$ is a phenomenological parameter of the model and serves to quantify the momentum dependence of the interaction. It is known to be $\gamma=1$ for self-interacting scalar fields with a $\lambda \phi^4$ potential, it is $\gamma=0$ for constant cross sections. It has been argued that in QCD effective kinetic theories $\gamma=1/2$ \cite{Dusling:2009df}. One can vary $\gamma$ and investigate different functional forms which are more phenomenologically relevant. However, negative values of $\gamma$ may lead to infrared divergences of the momentum integrals when the mass of the particles vanishes (which is the case in our chiral models, since the masses vanish at very high temperatures). Therefore, we present results for a few positive values of $\gamma$.

\section{Chapman-Enskog expansion and transport coefficients}
\label{chapmanexpansionsection}

In this Section we briefly review the first-order correction to the distribution function $f^{(0)}$ using the Chapman-Enskog expansion \cite{chap1, chap2, chap3, Denicol-rischke_2022, groot} of the relativistic Boltzmann equation with a thermal mass (for details, see Ref. \cite{Rocha:2022fqz}). With the first correction to the distribution function, it is possible to compute the transport coefficients. 

The Chapman-Enskog expansion consists in obtaining a perturbative solution of the relativistic Boltzmann equation where the single-particle momentum distribution function is expanded in a series of powers of spacelike gradients. This is done inserting a book-keeping parameter in all terms of the relativistic Boltzmann equation that contain a derivative, as in
\begin{equation}
\epsilon (p^{\mu}\partial_{x^{\mu}}f_{\textbf{p}}+\frac{1}{2}\partial_{i}M^{2}(T)\partial^{i}_{(\mathbf{p})}f_{\mathbf{p}})= f_{0\mathbf{p}}\hat{L}\phi_{\mathbf{p}} \,.
\end{equation}
The asymptotic behavior of the relativistic Boltzmann equation is recovered by truncating the series and setting $\epsilon=1$.

The zeroth-order solution to this expansion leads to the equilibrium distribution function, $f_{\mathbf{p}}=f^{(0)}_{\mathbf{p}}$. The first-order solution is obtained solving the following equation for $\phi^{(1)}_{\mathbf{p}}=f^{(1)}_{\mathbf{p}}/f^{(0)}_{\mathbf{p}}$:
\begin{equation}
\begin{aligned}
\label{boltzmann}
&\left(A_{p}\theta-\beta p^{\langle\mu}p^{\nu\rangle}\sigma_{\mu\nu}\right)f_{\mathbf{0p}}\\
&=
-\frac{E_{\mathbf{p}}}{\tau_{R}}f_{\mathbf{0p}}\Biggl [\phi_{\mathbf{p}}^{(1)}-\\
&
\frac{\left\langle\frac{\phi_{\mathbf{p}}^{(1)}E_{\mathbf{p}}^{2}}{\tau_{R}}\right\rangle_{0}}{\left\langle\frac{E_{\mathbf{p}}^{3}}{\tau_{R}}\right\rangle_{0}}E_{\mathbf{p}}-\frac{\left\langle\frac{\phi_{\mathbf{p}}^{(1)}E_{\mathbf{p}}p^{\langle\mu\rangle}}{\tau_{R}}\right\rangle_{0}}{\frac{1}{3}\left\langle\Delta^{\alpha\beta}p_{\alpha}p_{\beta}\frac{E_{\mathbf{p}}}{\tau_{R}}\right\rangle_{0}}p_{\langle\mu\rangle}\Biggl] \,,
\end{aligned}
\end{equation}
where we defined the rank-2 irreducible projector $p^{\langle\mu}p^{\nu\rangle}=\Delta^{\mu\nu\alpha\beta}p_{\alpha}p_{\beta}$, which makes use of the traceless and double-symmetric tensor 
\begin{equation}
\Delta^{\mu\nu\alpha\beta}=\frac{1}{2}(\Delta^{\mu\alpha}\Delta^{\nu\beta}+\Delta^{\mu\beta}\Delta^{\nu\alpha})-\frac{1}{3}\Delta^{\mu\nu}\Delta^{\alpha\beta} \,. 
\end{equation}
The solution for $\phi^{(1)}$ was computed in Ref. \cite{Rocha:2022fqz}, and is given by
\begin{equation}
\label{phi1}
\phi^{(1)}_{\mathbf{p}}=F^{(0)}_{\mathbf{p}}\theta+F^{(2)}_{\mathbf{p}}p^{\langle\mu}p^{\nu\rangle}\sigma_{\mu\nu} \,,
\end{equation}
where we defined the following scalar functions of $E_{\mathbf{p}}$
\begin{equation}
F^{(0)}_{\mathbf{p}}=-\frac{\tau_{R}}{E_{\mathbf{p}}}A_{\mathbf{p}}+\frac{1}{I_{1,0}}\left\langle-\frac{\tau_{R}}{E_{\mathbf{p}}}A_{\mathbf{p}}\right\rangle_{0}E_{\mathbf{p}} \,,
\end{equation}
\begin{equation}
F^{(2)}_{\mathbf{p}}=\beta\frac{\tau_{R}}{E_{\mathbf{p}}} \,.
\end{equation}

The transport coefficients are then obtained identifying the following constitutive relations \cite{Denicol-rischke_2022},
\begin{equation}
\Pi \approx -\frac{1}{3}\left\langle \left(\Delta^{\mu\nu}p_{\mu}p_{\nu}\right)\phi_{
p}^{(1)}\right\rangle_{0}=-\zeta \theta \,,
\end{equation}
\begin{equation}
\delta \epsilon \approx \left \langle E^{2}_{p}\phi_{p}^{(1)}\right\rangle_{0}=\chi \theta \,,
\end{equation}
\begin{equation}
\pi^{\mu\nu} \approx \left\langle \phi_{p}^{(1)}\right\rangle_{0}p^{\langle\mu}p^{\nu\rangle} =2\eta\sigma^{\mu\nu}\,,
\end{equation}
with $\Pi$ being the bulk viscous pressure, $\delta \epsilon$ is the nonequilibrium correction to the energy density, and $\pi^{\mu\nu}$ is the shear-stress tensor. We further introduced the expansion rate  $\theta = \partial_\mu u^\mu$ and the shear tensor $ \sigma^{\mu\nu} = \Delta^{\mu\nu\alpha\beta} \partial_\alpha u_\beta$, with $u^\mu$ being the four-velocity of the fluid. Here $\zeta$, $\chi$ and $\eta$ are transport coefficients and stand for the bulk viscosity, the energy deviation and the shear viscosity, respectively. 
Using the first-order solution of the Chapman-Enskog expansion, Eq. (\ref{phi1}), the transport coefficients are given by the following microscopic expressions,
\begin{equation}
\zeta=-\frac{1}{3}\left\langle \left(\Delta^{\mu\nu}p_{\mu}p_{\nu}\right)A_{\mathbf{p}}\frac{\tau_{R}}{E_{\mathbf{p}}}\right\rangle_{0}-\left\langle\frac{\tau_{R}}{E_{\mathbf{p}}} A_{\mathbf{p}}\right\rangle_{0} \frac{I_{3,1}}{I_{1,0}} \,,
\end{equation}
\begin{equation}
\eta=\frac{\beta}{15}\left\langle \left(\Delta^{\mu\nu}p_{\mu}p_{\nu}\right)^{2}\frac{\tau_{R}}{E_{\mathbf{p}}}\right\rangle_{0}\,,
\end{equation}
\begin{equation}
\chi=\left\langle\frac{\tau_{R}}{E_{\mathbf{p}}} A_{\mathbf{p}}\right\rangle_{0} \frac{I_{3,0}}{I_{1,0}}-\langle A_{p}\tau_{R}E_{\mathbf{p}}\rangle_{0}\,.
\end{equation}

One can also show that the entropy production is given by \cite{Rocha:2022fqz}
\begin{equation}
\partial_{\mu}S^{\mu}\approx \zeta_{s}\theta^{2}+2\eta\sigma^{\mu\nu}\sigma_{\mu\nu} \,,
\end{equation}
where the new transport coefficient $\zeta_{s}$ that emerges is related to the previous ones via
\begin{equation}
\zeta_{s}=\zeta+c_{s}^{2}\chi \,.
\end{equation}
This transport coefficient is independent of the matching condition \cite{Rocha:2022fqz}. In the Landau frame, there is no energy diffusion by definition, thus $\chi=0$, which implies that $\zeta_{s}=\zeta$. In other words, $\zeta_{s}$ would be identical to the bulk viscosity had we used the Landau frame.

\section{Discussion of results}
\label{discussionsection}

Let us now discuss and compare the transport coefficients obtained using the LSMq and NJL models to generate the thermal mass that enters as an input in the kinetic theory, as discussed previously.

\begin{figure}[t]
\centering
\includegraphics[width=.48\textwidth]{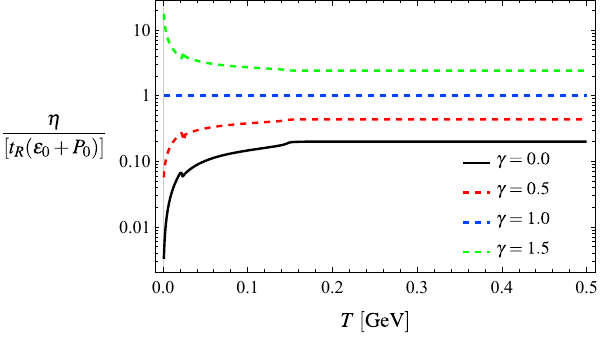}
    \caption{Normalized shear viscosity as a function of the temperature in the LSMq model for several values of the phenomenological parameter $\gamma$.}
\label{shearlinear}
\end{figure}

\begin{figure}[t]
\centering
    \includegraphics[width=.48\textwidth]{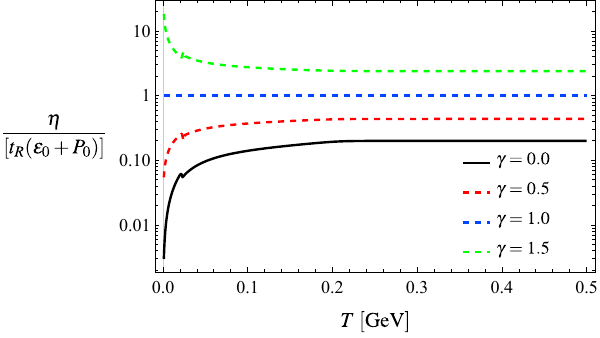}
    
    \caption{Normalized shear viscosity as a function of the temperature in the NJL model for several values of the phenomenological parameter $\gamma$.}
    \label{shearnjl}
\end{figure}

Figures \ref{shearlinear} and \ref{shearnjl} show that both models produce similar results for the shear viscosity. Chiral symmetry restoration does not bring about any qualitative feature beyond maintaining the shear viscosity at a constant value for temperatures higher than the critical temperature. The reason is that, in the limit $M(T)/T \rightarrow 0$, the asymptotic behavior for the shear viscosity is mass independent \cite{Rocha:2022fqz}
\begin{equation}
\label{eq:eta_assymptotic}
\frac{\eta}{t_{R}(\epsilon_{0}+P_{0})}\sim\frac{\Gamma(\gamma+5)}{120},
\end{equation}
where $\Gamma(z)$ denotes the gamma Function \cite{gradshteyn2007}.

\begin{figure}[h]
    \includegraphics[width=.48\textwidth]{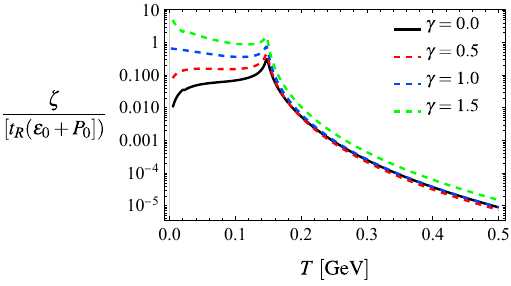}
    \caption{Normalized bulk viscosity as a function of the temperature in the LSMq model for several values of the phenomenological parameter $\gamma$.}
    \label{bulkLinear}
\end{figure}

\begin{figure}[h]
\includegraphics[width=.48\textwidth]{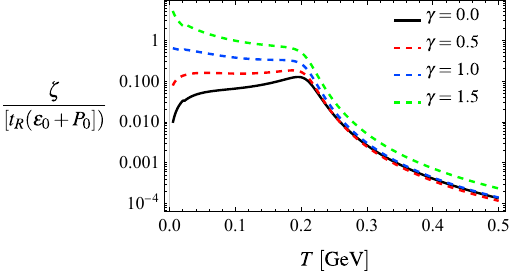}
    \caption{Normalized bulk viscosity as a function of the temperature in the NJL model for several values of the phenomenological parameter $\gamma$.}
    \label{bulknjl}
\end{figure}

\begin{figure}[h]
\includegraphics[width=.48\textwidth]{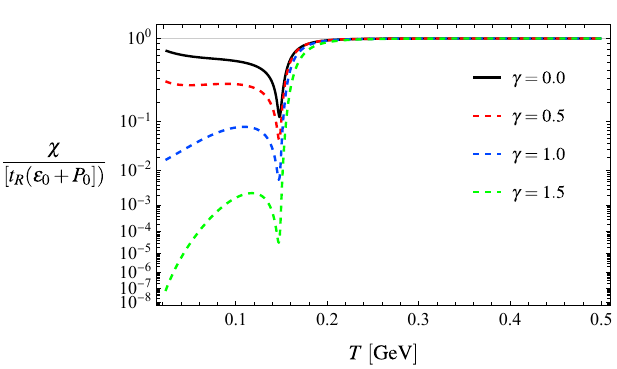}
    \caption{Normalized $\chi$ as a function of the temperature in the LSMq model for several values of the phenomenological parameter $\gamma$.}
    \label{deviation}
\end{figure}

\begin{figure}[h]
    \includegraphics[width=.48\textwidth]{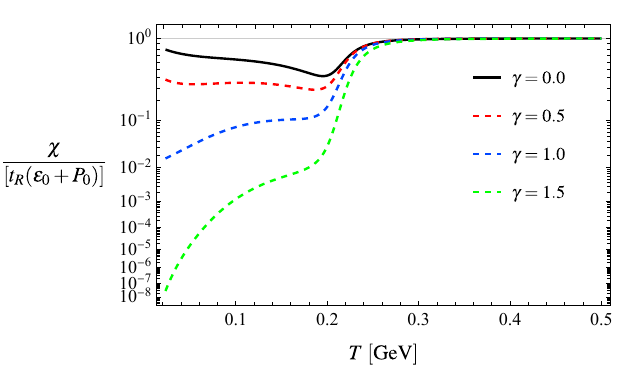}
    \caption{Normalized $\chi$ as a function of the temperature in the NJL model for several values of the phenomenological parameter $\gamma$.}
    \label{deviationnjl}
\end{figure}

The bulk viscosity and the coefficient of energy deviation are shown for both models in Figs.~\ref{bulkLinear}, \ref{bulknjl}, \ref{deviation} and \ref{deviationnjl}. One can see that the approximate chiral symmetry restoration strongly reduces both transport coefficients. The behavior of these transport coefficients in the limit $M(T)/T \rightarrow 0$ is given by the following relations \cite{Rocha:2022fqz}
\begin{equation}
\begin{aligned}
&\frac{\zeta}{t_{R}(\epsilon_{0}+P_{0})}\sim-\frac{1}{864} M(T) \frac{d}{dT}\left(\frac{M(T)}{T}\right)\\
&\times(\gamma^{4}+10\gamma^{3}+11\gamma^{2}-22\gamma+120)\Gamma(\gamma+1) \,,
\end{aligned}
\end{equation}
\begin{equation}
\begin{aligned}
&\frac{\chi}{t_{R}(\epsilon_{0}+P_{0})}\sim\frac{1}{288} M(T) \frac{d}{dT}\left(\frac{M(T)}{T}\right)\\
&\times(\gamma^{4}+10\gamma^{3}+11\gamma^{2}-22\gamma+120)\Gamma(\gamma+1) \,.
\end{aligned}
\end{equation}

Since the normalized bulk viscosity and the normalized coefficient of energy deviation are proportional to the thermal mass, both transport coefficients go to zero in the chiral limit. Figs.~\ref{bulkLinear} and \ref{deviation} show that chiral restoration in the LSMq model has a more prominent effect in these transport coefficients since it happens faster within temperature range than in the NJL model.

\begin{figure}[h]
    \includegraphics[width=.48\textwidth]{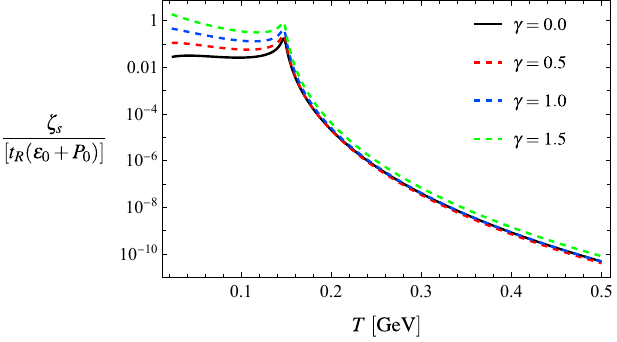}
    \caption{Normalized $\zeta_{s}$ as a function of the temperature in the LSMq model for several values of the phenomenological parameter $\gamma$.}
    \label{bulkslinear}
\end{figure}

\begin{figure}[h!]
    \includegraphics[width=.48\textwidth]{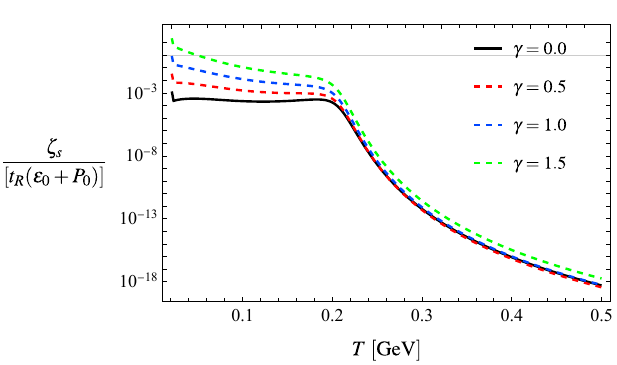}
   
    \caption{Normalized $\zeta_{s}$ as a function of the temperature in the NJL model for several values of the phenomenological parameter $\gamma$.}
    \label{bulksnjl}
\end{figure}

\begin{figure}[h!]
    \includegraphics[width=.48\textwidth]{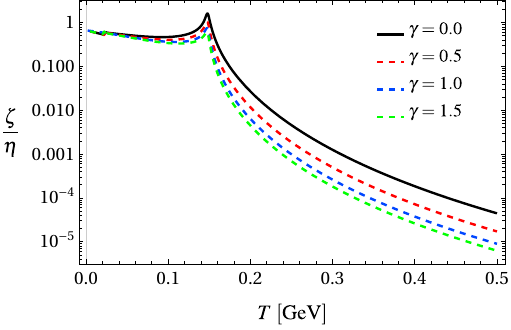}
   
    \caption{Bulk-shear ratio as a function of temperature in the LSMq model for several values of phenomenological parameter $\gamma$}
    \label{zetaeta}
\end{figure}

Figures \ref{bulkslinear} and \ref{bulksnjl} show that $\zeta_{s}$ is way smaller than $\zeta$ and goes down very fast after the chiral symmetry restoration critical temperature. The reason is essentially the same as before, with the asymptotic behavior given by \cite{Rocha:2022fqz} 
\begin{equation}
\begin{aligned}
&\frac{\zeta_{s}}{t_{R}(\epsilon_{0}+P_{0})}\sim\frac{1}{10368} M(T)^{2} \left[\frac{d}{dT}\left(\frac{M(T)}{T}\right)\right]\\&\times(\gamma^{4}+10\gamma^{3}+11\gamma^{2}-22\gamma+120)\Gamma(\gamma+1) \,.
\end{aligned}
\end{equation}
As before, the LSMq model exhibits a stronger effect, leading to a more radical decrease in the normalized $\zeta_{s}$. Moreover, $\zeta_{s}$  increases abruptly right before the critical temperature. This shows that a faster chiral restoration leads to a faster reduction in bulk viscosity, as well as an abrupt increase before the critical temperature. This difference in behavior is also clearly seen in the coefficient of energy deviation (cf. Figs. \ref{deviation} and \ref{deviationnjl}).

\begin{figure}[h!]
    \includegraphics[width=.48\textwidth]{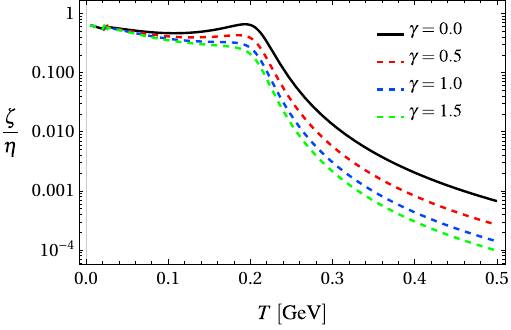}
   
    \caption{Bulk-shear ratio as a function of temperature in the NJL model for several values of phenomenological parameter $\gamma$}
    \label{zetaetanjl}
\end{figure}

Figures \ref{zetaeta} and \ref{zetaetanjl} display the ratio of the bulk viscosity to the shear viscosity, $\zeta/\eta$, as a function of the temperature, for both models. As seen, the same behavior in the chiral restoration region is exhibited. This ratio in temperatures above the critical temperature becomes smaller as $\gamma$ is increased. This can be explained due to the asymptotic behavior of the shear viscosity in the limit $M(T)/T \rightarrow 0$ (\ref{eq:eta_assymptotic}) as already seen in Figs. \ref{shearlinear} and \ref{shearnjl}. At small temperatures (below the critical temperature) the effect of the parameter $\gamma$ is small.

We should remark that 
the transport coefficients obtained here differ, with the exception of the shear viscosity, from the ones obtained using the trace anomaly from lattice simulations as done in Ref. \cite{Rocha:2022fqz}. In any case, one does not expect them to be similar, since the LSMq model and NJL model are low-energy effective descriptions and have different thermal quark masses.

\section{Summary and outlook}
\label{conclusion}

We computed transport coefficients using a quasiparticle description in which the mass of the particles depend on the temperature and are extracted from chiral models. We adopted the linear sigma model coupled with constituent quarks and NJL model. This produces a temperature-dependent background field in the corresponding relativistic Boltzmann equation. We then used a Chapman-Enskog expansion and a relaxation-time approximation to calculate the transport coefficients.

We showed that chiral symmetry restoration produces a strong decrease of the bulk viscosity and the coefficient of energy deviation after the critical temperature, and that the shear viscosity reaches a constant value. Right before chiral restoration, the bulk viscosity increases and the energy deviation coefficient decreases. Chiral restoration also decreases the speed of sound around the critical temperature which afterward keeps increasing until reaching the conformal limit. The linear sigma model coupled with constituent quarks exhibits a more pronounced effect compared to the NJL model as a consequence of chiral restoration in the former being faster as a function of temperature than in the latter. 

We expect the inclusion of Polyakov loops, as commonly done in extensions of the chiral models adopted in this paper, would make the chiral transition smoother. If that is the case, this might lead to a milder temperature dependence of the bulk viscosity and speed of sound near the phase transition.

There are several possible directions for improvement of our description. For instance, a more immediate follow up of this work would be to calculate the second-order transport coefficients, allowing for fluid-dynamical simulations of these type of systems. A more nontrivial application that we wish to pursue is the inclusion of the dynamics of the chiral field, which will lead to nonequilibrium corrections to the thermal mass.

\acknowledgments

This work was partially supported by CAPES (Finance Code 001), Conselho Nacional de Desenvolvimento Cient\'{\i}fico e Tecnol\'{o}gico (CNPq), Funda\c c\~ao Carlos Chagas Filho de Amparo \` a Pesquisa do Estado do Rio de Janeiro (FAPERJ), and INCT-FNA (Process No. 464898/2014-5).

\section*{DATA AVAILABILITY}
No data were created or analyzed in this study.


\bibliography{references}
\end{document}